\documentclass[namedreferences]{solarphysics}
\usepackage[hyperref,optionalrh]{spr-sola-addons}
\usepackage{graphicx}
\usepackage{amssymb}            
\usepackage{color}                                        
\usepackage{breakurl}

\newcommand{\arcsec}{^{\prime\prime}}

\chardef\us=`\_

\begin{document}

\begin{article}
\begin{opening}
 
\title{Transverse Coronal-Loop Oscillations Induced by the Non-radial Eruption of a Magnetic Flux Rope}

\author[addressref={aff1,aff2},corref,email={zhangqm@pmo.ac.cn}]{\inits{Q.M.}\fnm{Q.M.}~\lnm{Zhang}\orcid{0000-0003-4078-2265}}
\author[addressref={aff1,aff2},corref]{\inits{J.L. Chen}\fnm{J.L.}~\lnm{Chen}\orcid{0000-0002-2436-0516}}
\author[addressref={aff1,aff2},corref]{\inits{S.T. Li}\fnm{S.T.}~\lnm{Li}}
\author[addressref={aff1,aff2},corref]{\inits{L. Lu}\fnm{L.}~\lnm{Lu}\orcid{0000-0002-3032-6066}}
\author[addressref={aff1,aff2},corref]{\inits{D. Li}\fnm{D.}~\lnm{Li}\orcid{0000-0002-4538-9350}}
\address[id=aff1]{Key Laboratory of Dark Matter and Space Astronomy, Purple Mountain Observatory, CAS, Nanjing 210023, China}
\address[id=aff2]{School of Astronomy and Space Science, University of Science and Technology of China, Hefei 230026, China}

\runningauthor{Q. M. Zhang et al.}
\runningtitle{Transverse coronal-loop oscillations}

\begin{abstract}
We investigate the transverse coronal-loop oscillations induced by the eruption of a prominence-carrying flux rope on 7 December 2012. 
The flux rope originating from NOAA Active Region (AR) 11621 was observed in extreme-ultraviolet (EUV) wavelengths by the \textit{Atmospheric Imaging Assembly} (AIA) 
onboard the \textit{Solar Dynamics Observatory} (SDO) spacecraft and in H$\alpha$ line center by the ground-based telescope at the \textit{Big Bear Solar Observatory} (BBSO).
The early evolution of the flux rope is divided into two steps: a slow rise phase at a speed of $\approx$230\,km\,s$^{-1}$ and a fast rise phase at a speed of $\approx$706\,km\,s$^{-1}$.
The eruption generates a C5.8 flare and the onset of the fast rise is consistent with the hard X-ray (HXR) peak time of the flare. 
The embedded prominence has a lower speed of $\approx$452\,km\,s$^{-1}$.
The eruption is significantly inclined from the local solar normal by $\approx$60$^{\circ}$, suggesting a typical non-radial eruption.
During the early eruption of the flux rope, the nearby coronal loops are disturbed and experience independent kink-mode oscillations in the horizontal and vertical directions.
The oscillation in the horizontal direction has an initial amplitude of $\approx$3.1\,Mm, a period of $\approx$294\,seconds, and a damping time of $\approx$645\,seconds. 
It is most striking in 171\,{\AA} and lasts for three to four cycles. The oscillations in the vertical directions are observed mainly in 171, 193, and 211\,{\AA}.
The initial amplitudes lie in the range of 3.4\,--\,5.2\,Mm, with an average value of 4.5\,Mm. The periods are between 407\,seconds and 441\,seconds, with an average value of 423\,seconds.
The oscillations are damping and last for nearly four cycles. The damping times lie in the range of 570\,--\,1012\,seconds, with an average value of 741\,seconds.
Assuming a semi-circular shape of the vertically oscillating loops, we calculate the loop lengths according to their heights. Using the observed periods, we carry out coronal seismology 
and estimate the internal Alfv\'{e}n speeds (988\,--\,1145\,km\,s$^{-1}$) and the magnetic-field strengths (12\,--\,43\,G) of the oscillating loops.
\end{abstract}
\keywords{Prominences, Active; Coronal Mass Ejections, Initiation and Propagation; Magnetic fields, Corona; Waves, Magnetohydrodynamic}
\end{opening}

\section{Introduction}  \label{s-intro}
A magnetic flux rope is a bundle of twisted field lines that wind around their common axis \citep[see][and references therein]{liu20}. 
The accumulated twist number [$\mathcal{T}_w$] can reach up to one to three turns \citep{liu16,guo21a}.
Flux ropes play an essential role in the genesis of solar eruptions \citep{chen17,cheng17}, including prominence eruptions \citep{rust96,ama14}, flares \citep{td99,jan15,wang15}, 
and coronal mass ejections \citep[CMEs:][]{dere99,vou13,pats20}.
It is still controversial whether flux ropes are generated before eruptions \citep{can09,gre09,zhang12,zqm17,jam18,yan18,chen19,he20,nin20} or during eruptions \citep{cheng11,gou19}.
Photospheric flux cancellation is found to be important in the formation of flux ropes before eruptions \citep{gre11,sav12}, 
while tether-cutting magnetic reconnection in the corona is believed to be an effective mechanism of flux-rope formation during eruptions \citep{jos14,xue17}.
Flux ropes are frequently observed to have very high temperatures ($\approx$10\,MK), which are best revealed in extreme-ultraviolet (EUV) 94\,{\AA} and 131\,{\AA} 
and therefore termed ``hot channel'' \citep{cheng11,cheng12,cheng13,cheng14}. \citet{nin15} analyzed 141 M-class and X-class flares.
About 32\,\% of the events are associated with hot channels and almost half of the eruptive events are related to a hot channel configuration.

The directions of prominence eruptions and associated CMEs are not always radial. 
\citet{mc15} investigated the properties of 904 prominence and filament eruptions observed by the \textit{Solar Dynamics Observatory} (SDO) in detail. 
It is found that the percentage of non-radial eruptions reaches 12\,\%.
\citet{devi21} reported a non-radial prominence eruption away from the local vertical 
with an inclination angle of $\gamma=48^{\circ}$, which is attributed to the easier channel provided by the open and high-lying magnetic field.
Using stereoscopic observations from the twin spacecraft of the \textit{Solar TErrestrial RElations Observatory} \citep[STEREO:][]{kai08},
\citet{gos09} observed a partial filament eruption that was highly inclined to the solar normal with an inclination angle of $\gamma=47^{\circ}$, which is close to that reported by \citet{wil05}.
Combining the observations from the STEREO-\textit{Ahead} (hereafter STA) and the \textit{Atmospheric Imaging Assembly} \citep[AIA:][]{lem12} onboard SDO,
\citet{sun12b} reported a non-radial, jet-like eruption following a markedly inclined trajectory with $\gamma=66^{\circ}$.
Combining the observations from STEREO-\textit{Behind} (hereafter STB) and SDO/AIA, \citet{bi13} investigated the rotation and non-radial propagation of a filament, the later of which resulted from 
interaction between the filament eruption and the overlying pseudo-streamer. In an extreme case, a nearly 90$^\circ$ deflected filament eruption and the related CME were noticed by \citet{yang18}.
State-of-the-art numerical simulations indicate that the imbalance of the bipole leads to a negative magnetic pressure gradient in the $x$-direction, 
which prevents the flux rope from expanding symmetrically \citep{au10,kli13,in18,jia18}.
\citet{guo21b} performed a magnetohydrodynamic (MHD) simulation of a C7.7 class flare, which was generated by a non-radial prominence eruption on 21 June 2011 \citep{zhou17}.

Solar eruptions can potentially generate kink-mode, transverse oscillations of the adjacent coronal loops \citep{asch99,naka99,zim15} and quasi-periodic pulsations \citep[QPP:][]{zim21}. 
The polarization of the transverse loop oscillations could be horizontal \citep{wht12a,nis13,li17,li18,zqm15,zqm20b,zqm20a,dai21} 
or vertical \citep{wang04,gos12,wht12b,sim13,sri13,kim14,dud16,ver17,ree20}.
In most cases, the amplitudes of kink oscillations decay with time as a result of resonant absorption, phase mixing, wave leakage, 
or Kelvin--Helmholtz instability \citep{go02,ofm02,rud02,ter06,god16b,ant17,nech19}. The damping time [$\tau$] is roughly proportional to the period [$P$] \citep{ver13}, 
and the quality factor [$\frac{\tau}{P}$] has a power-law dependence on the amplitude with the exponent of -0.5 \citep{god16a}.
One of the applications of coronal seismology is the estimation of the coronal magnetic field and Alfv\'{e}n speed of the oscillating loops \citep{naka01,asch02,ver09,chen15}.
So far, there are few reports of transverse coronal-loop oscillations triggered by non-radial prominence eruptions.

On 7 December 2012, a prominence-carrying flux rope erupted from NOAA Active Region (AR) 11621 (N15W91) and propagated non-radially, producing a C5.8 flare and a fast CME.
Based on the revised cone model, \citet{zqm21} performed a 3D reconstruction of the CME simultaneously observed by SDO/AIA and STA/EUVI at 21:20:30 UT. 
The geometry and kinematics of the CME were derived. In this article, we report the simultaneous coronal-loop oscillations in the horizontal and vertical directions induced by the eruption in the same AR.
Data analysis is described in Section~\ref{s-obs}. The results are presented in Section~\ref{s-res} and compared with previous findings in Section~\ref{s-dis}. 
Finally, a brief summary is given in Section~\ref{s-sum}.

\begin{table}
\caption{Description of the observational parameters.}
\label{tab-1}
\tabcolsep 1.5mm
\begin{tabular}{lcccc}
  \hline
Instrument & $\lambda$   & Time &  Cadence & Pixel Size \\ 
                  & [{\AA}]         &  [UT] &  [s]           & [$\arcsec$] \\
  \hline
AIA & 131\,--\,211 & 21:10\,--\,21:50 &  12 & 0.6 \\
HMI & 6173       & 21:10\,--\,21:50 & 45 & 0.6 \\
EUVI & 195       & 21:05\,--\,23:05 & 300 & 1.6 \\
COR2 & WL      & 22:24\,--\,23:39 & 900 & 14.7 \\
LASCO/C2 & WL & 21:36\,--\,22:36 & 720 & 11.4 \\
BBSO      & 6563 & 21:00\,--\,21:50 & 60   &  1.1 \\
GOES     & 0.5\,--\,4 & 21:00\,--\,22:30 &  2.05 & ... \\
GOES     & 1\,--\,8      & 21:00\,--\,22:30  &  2.05 & ... \\
GBM       & 4\,--\,26\,keV & 21:05\,--\,21:30  &  0.256 & ... \\
  \hline
\end{tabular}
\end{table}

\section{Data Analysis} \label{s-obs}
The prominence eruption was tracked by the ground-based telescope at the \textit{Big Bear Solar Observatory} (BBSO) in the H$\alpha$ line center.
The eruption of the flux rope was detected by SDO/AIA in EUV wavelengths (131, 171, 193, and 211\,{\AA}).
The line-of-sight (LOS) magnetograms of the photosphere were observed by the \textit{Helioseismic and Magnetic Imager} \citep[HMI:][]{sch12} onboard SDO.
The AIA and HMI level\_1 data were calibrated using the standard Solar Software (SSW) program \textsf{aia\_prep.pro} and \textsf{hmi\_prep.pro}, respectively.
The H$\alpha$ and EUV images were coaligned using the cross-correlation method.

The eruption was also captured by the \textit{Extreme-Ultraviolet Imager} (EUVI) in the 
\textit{Sun Earth Connection Coronal and Heliospheric Investigation} \citep[SECCHI:][]{how08} package onboard STA,
which had a separation angle of $\approx$128$^{\circ}$ with respect to the Sun--Earth direction on 7 December 2012.
The CME driven by the flux-rope eruption was observed by the C2 white light (WL) coronagraph of the \textit{Large Angle Spectroscopic Coronalgraph} \citep[LASCO:][]{bru95} 
onboard the \textit{Solar and Heliospheric Observatory} (SOHO). 
The LASCO/C2 data were calibrated using the SSW program \textsf{c2\_calibrate.pro}. The CME was observed by the COR2 coronagraph onboard STA as well.
Calibrations of the COR2 and EUVI data were performed using the SSW program \textsf{secchi\_prep.pro}. The deviation of STA north--south direction from the solar rotation axis was corrected.
The soft X-ray (SXR) fluxes of the C5.8 flare were recorded by the \textit{Geostationary Operational Environmental Satellite} (GOES) spacecraft.
The hard X-ray (HXR) fluxes at different energy bands were obtained from the the \textit{Gamma-ray Burst Monitor} \citep[GBM:][]{mee09} onboard the \textit{Fermi} spacecraft.
The observational properties of the instruments are listed in Table~\ref{tab-1}.

\section{Results} \label{s-res}
\subsection{Eruption of the Prominence-carrying Magnetic Flux Rope} \label{s-fr}
\begin{figure} 
\centerline{\includegraphics[width=0.8\textwidth,clip=]{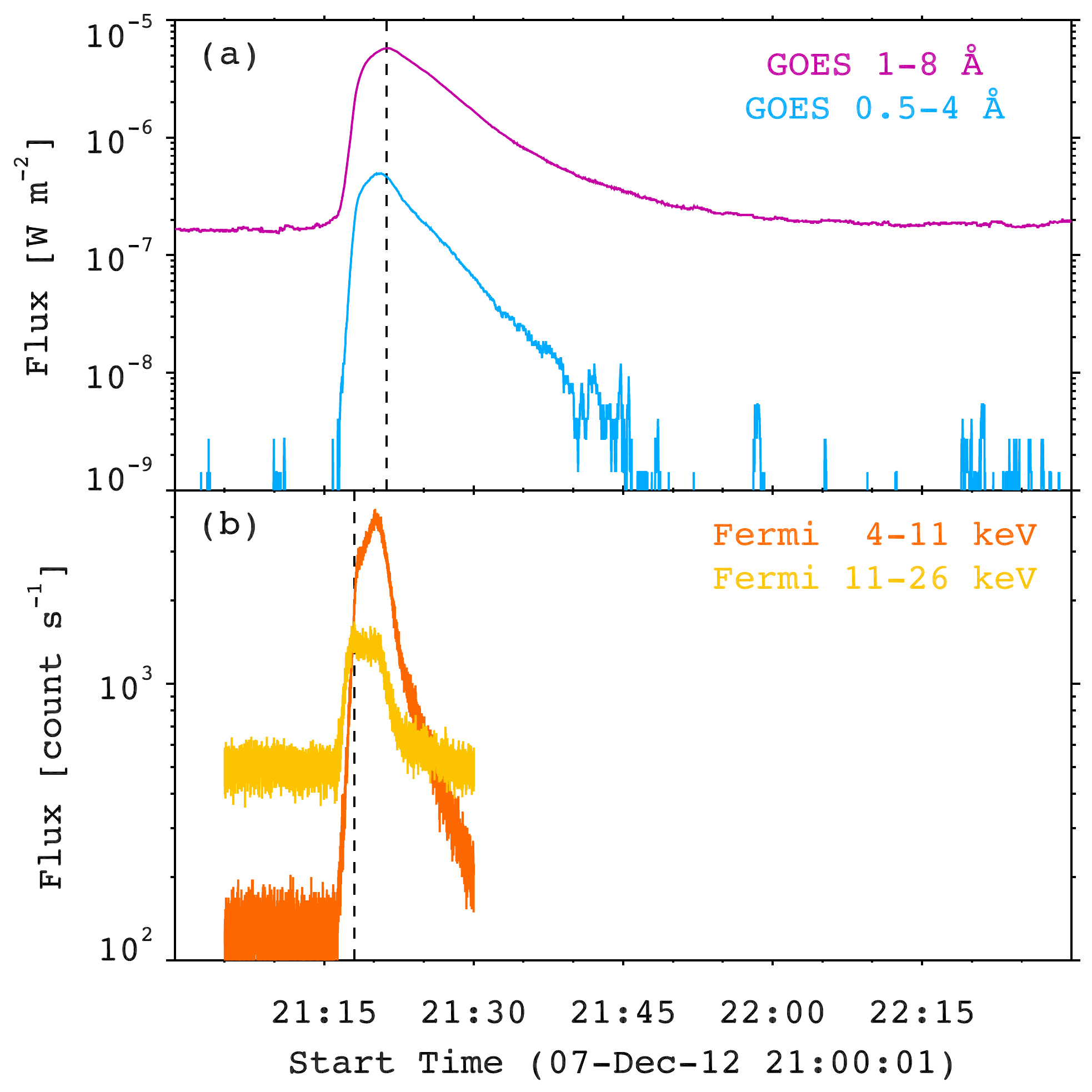}}
\caption{({\bf a}) Soft X-ray (SXR) light curves of the GOES C5.8 flare in 1\,--\,8\,{\AA} (\textit{red line}) and 0.5\,--\,4\,{\AA} (\textit{blue line}). 
The \textit{black-dashed line} denotes the peak time at 21:21:15 UT.
({\bf b}) Hard X-ray (HXR) light curves of the flare observed by \textit{Fermi}/GBM at 4\,--\,11\,keV (\textit{orange line}) and 11\,--\,26\,keV (\textit{yellow line}).
The \textit{black-dashed line} denotes the HXR peak time at 21:18:00 UT.}
\label{fig1}
\end{figure}

In Figure~\ref{fig1}, the upper panel shows the SXR light curves of the C5.8 flare in 1\,--\,8\,{\AA} (red line) and 0.5\,--\,4\,{\AA} (blue line).
It is clear that the SXR flux increases rapidly from $\approx$21:13:00 UT and reaches the peak value at $\approx$21:21:15 UT (black-dashed line) 
before declining gradually to the pre-flare level around 22:13:00 UT. Hence, the lifetime of the flare is $\approx$1 hr.
Figure~\ref{fig1}b shows the HXR light curves at 4\,--\,11\,keV (orange line) and 11\,--\,26\,keV (yellow line).
The black-dashed line denotes the HXR peak time at $\approx$21:18:00 UT when the rate of energy precipitation of flare-accelerated nonthermal electrons is maximum \citep{bro71}.

\begin{figure}
\centerline{\includegraphics[width=0.9\textwidth,clip=]{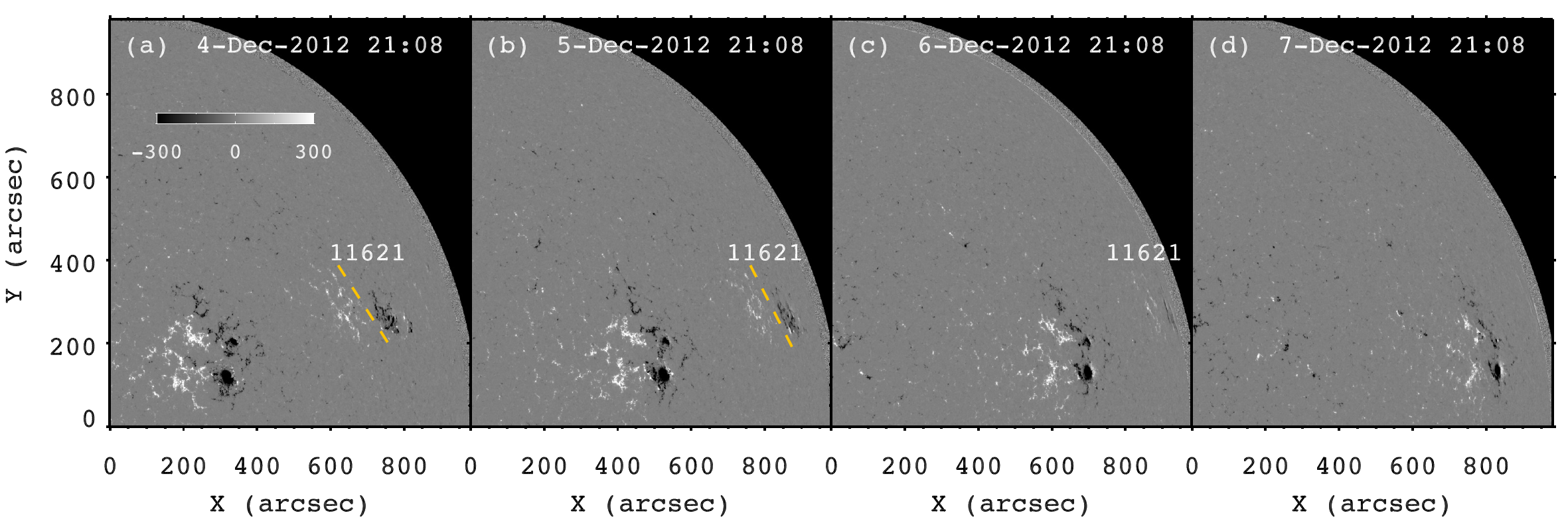}}
\caption{HMI LOS magetograms from 4 December to 7 December 2012.
AR 11621 hosting the C5.8 flare is marked. The \textit{gold-dashed lines} denote the approximate polarity inversion line (PIL).}
\label{fig2}
\end{figure}

Four selected LOS magetograms from 4 December to 7 December 2012 are displayed in Figure~\ref{fig2}. It is seen that AR 11621 was still distinguishable on 5 December.
As time goes by, it got blurred on 6 December close to the western limb and totally vanished on 7 December, implying that it had rotated to the backside of the Sun when the flare took place.

\begin{figure}
\centerline{\includegraphics[width=0.9\textwidth,clip=]{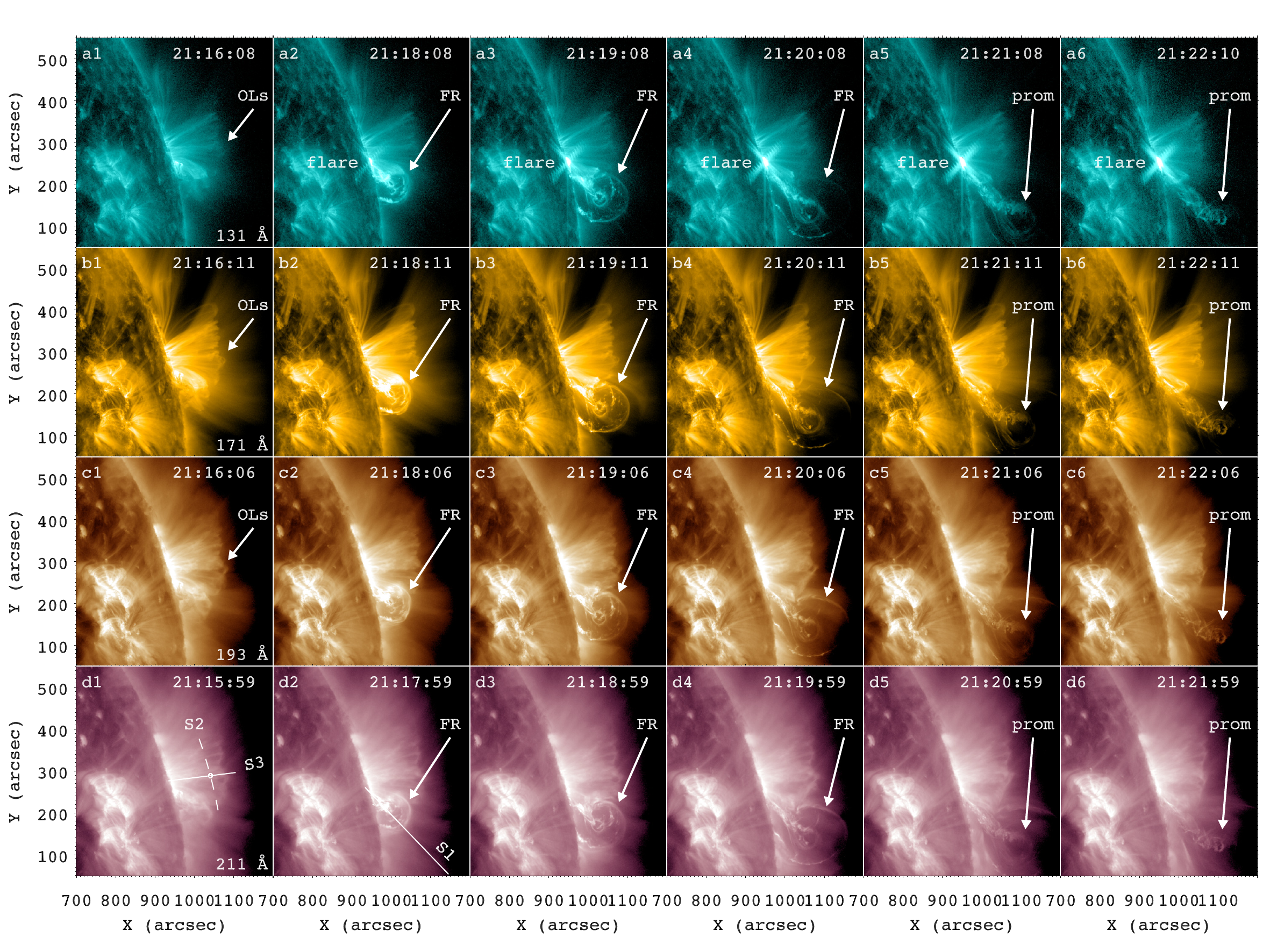}}
\caption{EUV images in 131\,{\AA} ({\bf a1\,--\,a6}), 171\,{\AA} ({\bf b1\,--\,b6}), 193\,{\AA} ({\bf c1\,--\,c6}), and 211\,{\AA} ({\bf d1\,--\,d6}).
The \textit{arrows} point to the oscillating loops (OLs), flux rope (FR), and eruptive prominence (prom).
In Panel {\bf d1}, two short slices (S2 and S3) are used to investigate the transverse oscillations of the coronal loops in the horizontal and vertical directions, respectively.
In Panel {\bf d2}, a long slice (S1) is used to investigate the evolution of the flux rope. An animation of this figure is available in the Electronic Supplementary Material (\textsf{animaia.mp4}).}
\label{fig3}
\end{figure}

The EUV images in Figure~\ref{fig3} illustrate the evolution of the prominence-carrying flux rope (see also the Electronic Supplementary Material \textsf{animaia.mp4}). 
As the flare occurs, the flux rope shows up before 21:18:00 UT (a2\,--\,d2). The bubble-like flux rope expands in size and propagates in the southwest direction. 
An embedded prominence follows the flux rope. It is obvious that the trajectory is severely inclined to the solar normal with $\gamma=60^{\circ}$, 
meaning that the event is a typical non-radial eruption \citep{zqm21}.
The thin leading edge of the flux rope could not be tracked after escaping the field-of-view (FOV) of AIA around 21:21:40 UT.
The eruption is evident in various EUV wavelengths, suggesting the multi-thermal nature of flux rope and prominence \citep{han13}.
We note that a group of coronal loops to the North of the flux rope are slightly disturbed during the eruption and oscillate for a long time, which will be described in Section~\ref{s-osc}.

\begin{figure} 
\centerline{\includegraphics[width=0.9\textwidth,clip=]{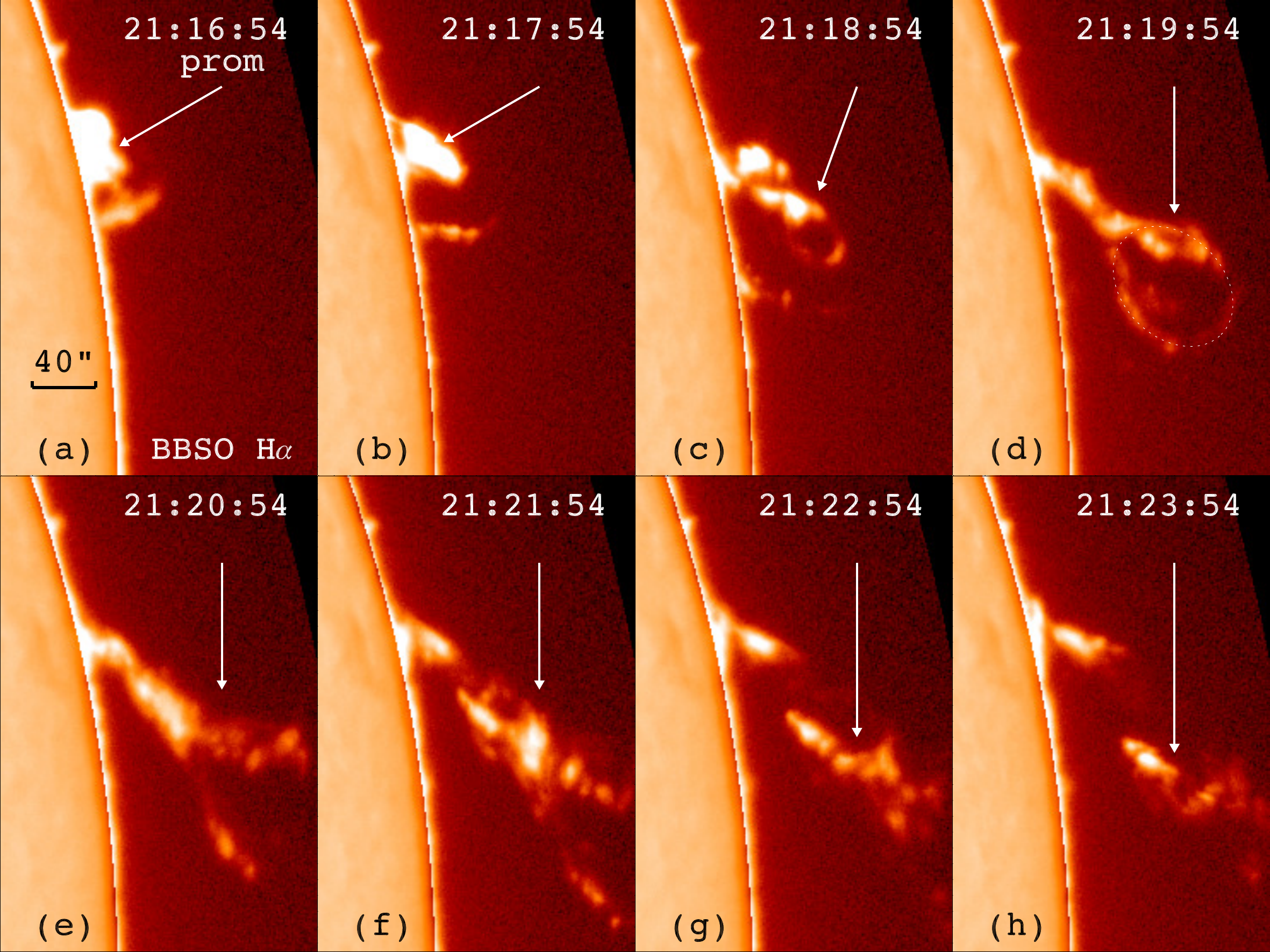}}
\caption{H$\alpha$ images observed by BBSO. The \textit{arrows} point to the eruptive prominence (prom).
In Panel {\bf d}, a dotted ellipse is used to fit the top segment of prominence.
An animation of this figure is available in the Electronic Supplementary Material (\textsf{animha.mp4}).}
\label{fig4}
\end{figure}

The prominence eruption is also vivid in H$\alpha$ line center (see animation in the Electronic Supplementary Material \textsf{animha.mp4}). 
Figure~\ref{fig4} shows the H$\alpha$ images observed by BBSO.
The prominence arises slowly at $\approx$21:16:00 UT and expands quickly, resembling a tennis racket (see Panel d).
The top segment is fitted with an ellipse (dotted line), whose major axis and minor axis have lengths of 86$\arcsec$ and 64$\arcsec$, respectively.
The location and shape of the prominence are consistent with that observed in EUV wavelengths (see Figure~\ref{fig3}a4\,--\,d4) 
and are in agreement with the typical $\mathsf{U}$-shaped prominence horns both in observation \citep{reg11} and numerical simulations \citep{xia14}.
The prominence continues ascending and the top segment escapes the FOV of BBSO. It is noticed that the bubble-like flux rope in EUV wavelengths is not distinct in H$\alpha$.

\begin{figure} 
\centerline{\includegraphics[width=0.9\textwidth,clip=]{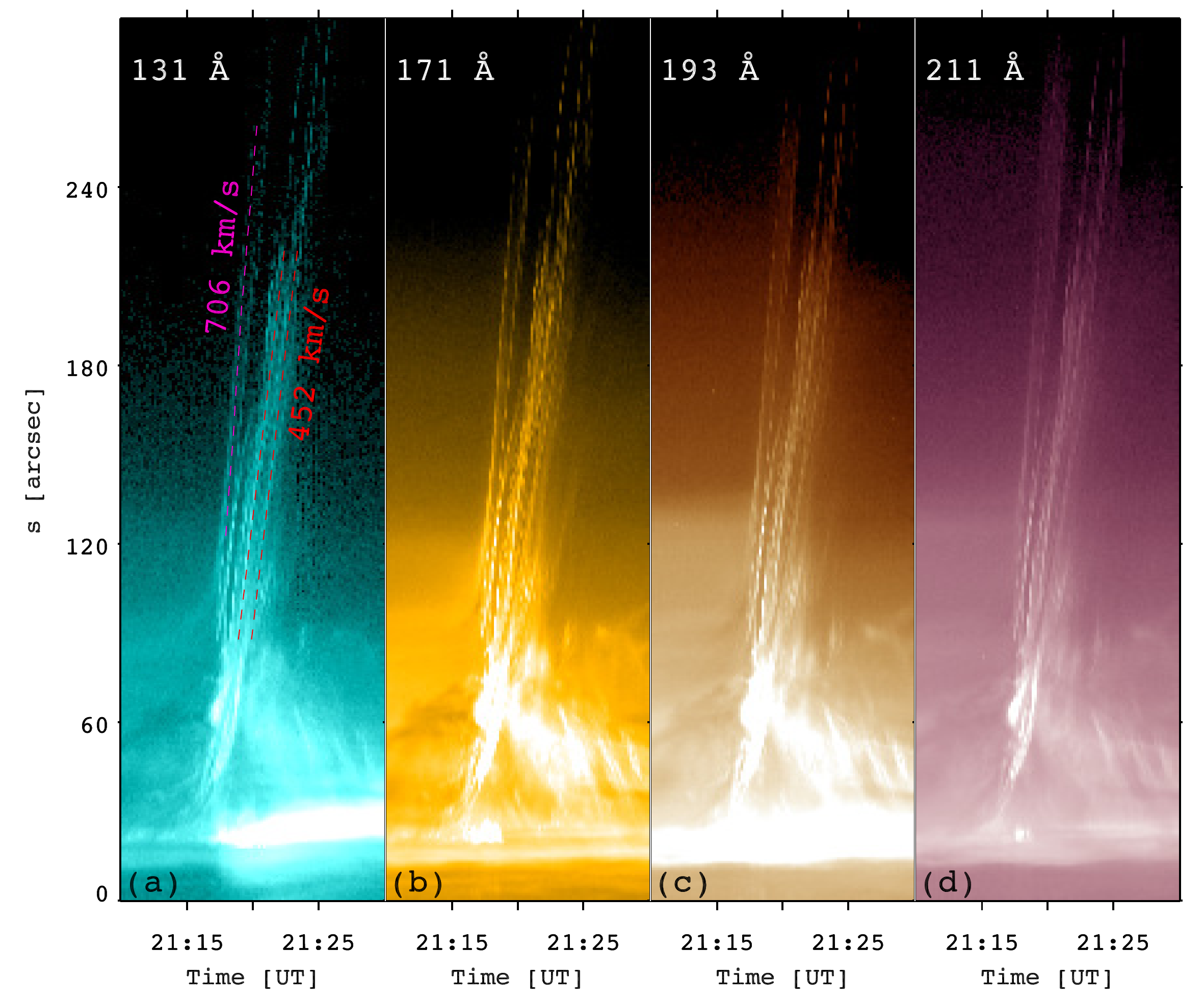}}
\caption{Time--distance diagrams of S1 in 131, 171, 193, and 211\,{\AA}.
The linear velocities of flux rope ($\approx$706\,km\,s$^{-1}$) and prominence ($\approx$452\,km\,s$^{-1}$) are indicated.}
\label{fig5}
\end{figure} 

To investigate the evolution of flux rope and prominence, we select a slice (S1) with a total length of $\approx$296$\arcsec$ along the direction of their propagation, which is shown in Figure~\ref{fig3}d2.
The time--distance diagrams of S1 in 131, 171, 193, and 211\,{\AA} are displayed in Figure~\ref{fig5}. 
In Figure~\ref{fig5}a, the positions of the flux rope leading edge are labeled with a magenta dashed line, 
whose slope represents its apparent linear speed ($\approx$706\,km\,s$^{-1}$). The positions of the following prominence are labeled with red-dashed lines, whose slopes represent its apparent linear
speed ($\approx$452\,km\,s$^{-1}$). Hence, the speed of the flux rope is $\approx$1.5 times larger than the prominence.

\begin{figure} 
\centerline{\includegraphics[width=0.9\textwidth,clip=]{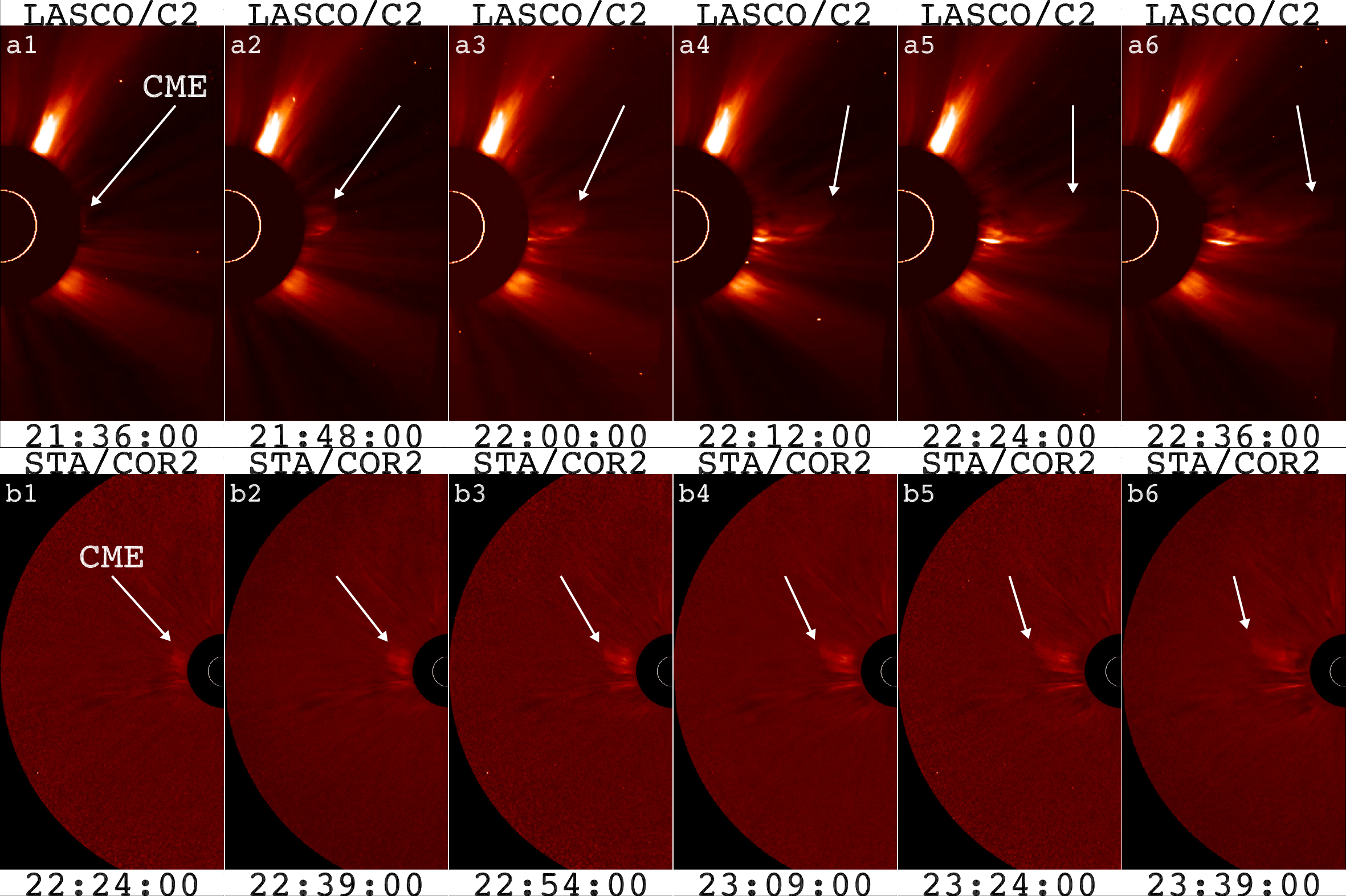}}
\caption{WL images of the CME observed by LASCO/C2 (\textit{upper panels}) and STA/COR2 (\textit{lower panels}).
The \textit{arrows} point to the CME leading front.}
\label{fig6}
\end{figure}

As mentioned in Section~\ref{s-intro}, the eruption of the flux rope drives a fast and wide CME.
In Figure~\ref{fig6}, the upper and lower panels show the WL images of the CME observed by LASCO/C2 and STA/COR2, respectively.
The leading fronts of the CME are pointed to by the arrows. The CME first appears at $\approx$21:36:00 UT in the LASCO/C2 FOV and propagates westward with a central position angle of 299$^{\circ}$.
The CME first appears at $\approx$22:24:00 UT in the STA/COR2 FOV and propagates eastward until $\approx$23:54:00 UT, when the CME becomes too weak to be identified.
It is shown that although the flux rope propagates non-radially in the FOV of AIA, the position angle of the related CME is close to the latitude of AR 11621, 
implying that the flux rope is probably influenced and redirected by the large-scale magnetic field.

\begin{figure}
\centerline{\includegraphics[width=0.9\textwidth,clip=]{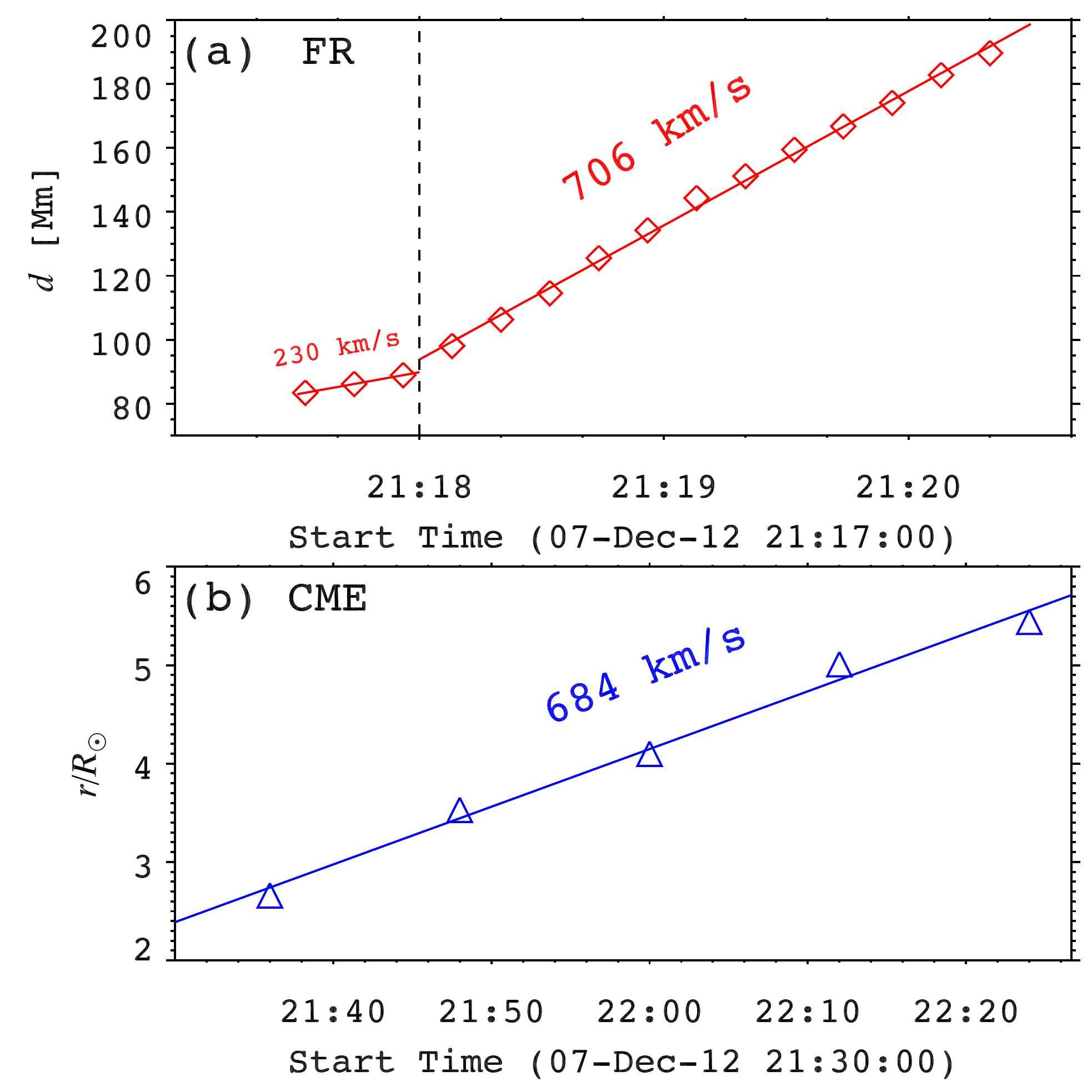}}
\caption{({\bf a}) Height variation of the flux-rope leading edge along S1 in 131\,{\AA}. 
The apparent linear speeds of slow rise ($\approx$230\,km\,s$^{-1}$) and fast rise ($\approx$706\,km\,s$^{-1}$) are indicated.
The \textit{black-dashed line} denotes the HXR peak time at 21:18:00 UT.
({\bf b}) Height variation of the CME leading front observed by LASCO/C2. The plane-of-sky linear speed ($\approx$684\,km\,s$^{-1}$) is indicated.}
\label{fig7}
\end{figure}

In Figure~\ref{fig7}, the upper panel shows the height evolution of the flux rope leading edge in 131\,{\AA}, which is determined manually in the direction of S1.
The movement of the flux rope is roughly divided into two phases by the HXR peak (black-dashed line): 
a slow rise ($\approx$230\,km\,s$^{-1}$) and a fast rise ($\approx$706\,km\,s$^{-1}$), which is in line with previous observations \citep{cheng13}.
The height evolution of the CME leading front in the FOV of LASCO/C2 is plotted in the lower panel of Figure~\ref{fig7}. The apparent speed ($\approx$684\,km\,s$^{-1}$) of CME is indicated. 
Taking the projection effect into account, the speeds of the flux rope and CME are comparable, validating that the flux rope serves as a driver of the CME \citep{cheng13}.

\begin{figure} 
\centerline{\includegraphics[width=0.9\textwidth,clip=]{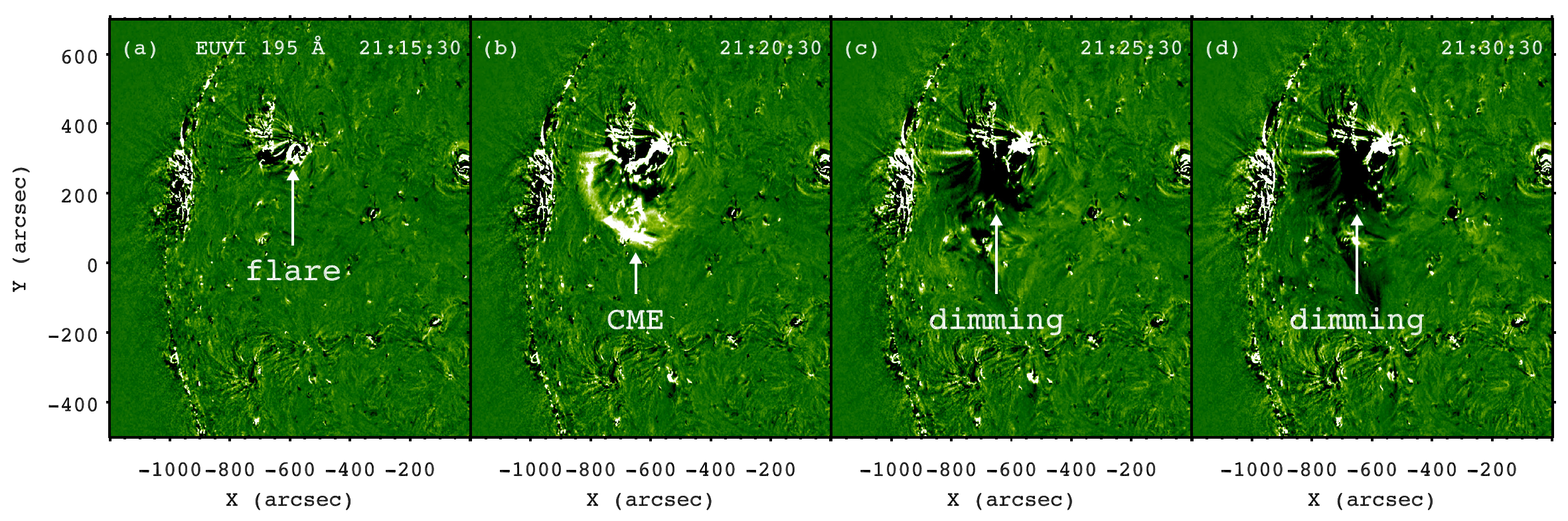}}
\caption{Base-difference images observed by STA/EUVI in 195\,{\AA}.
The \textit{arrows} point to the flare, CME front, and coronal dimming behind the CME.
An animation of this figure is available in the Electronic Supplementary Material (\textsf{anim195.mp4}).}
\label{fig8}
\end{figure}

The prominence eruption was observed by STA/EUVI from a different viewpoint (see animation in the Electronic Supplementary Material \textsf{anim195.mp4}). 
Four 195\,{\AA} base-difference images are displayed in Figure~\ref{fig8}. 
In Panel a, the source region of the flare and CME is located at (-590$\arcsec$, 300$\arcsec$) as indicated by the arrow. 
In Panel b, the arc-shaped CME leading front with enhanced intensity is pointed by the arrow. 
In Panels c\,--\,d, long-lasting coronal dimming behind the CME with reduced intensity and expanding area is the most striking feature \citep{tho98,zqm17}.
It is evident that as the CME propagates in the southeast direction, the dimming mainly extends in the south and east directions, rather than isotropically.

\subsection{Transverse Coronal-Loop Oscillations} \label{s-osc}
\begin{figure}
\centerline{\includegraphics[width=0.8\textwidth,clip=]{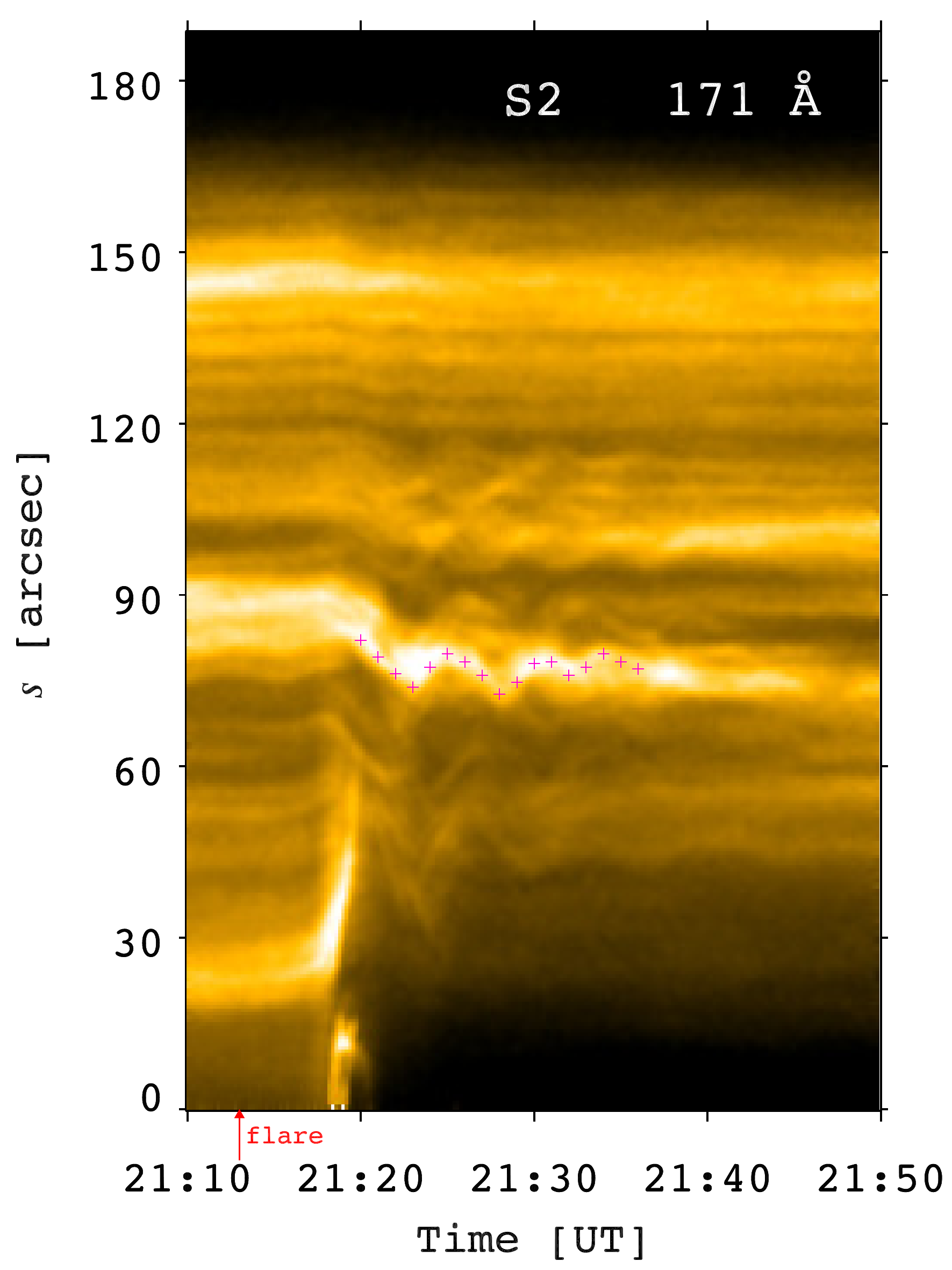}}
\caption{Time--distance diagram of S2 in 171\,{\AA}. The \textit{magenta-plus symbols} denote the positions of the oscillating loop with maximal EUV intensity.
The \textit{red arrow} on the $x$-axis indicates the starting time of flare. $s=0$ and $s=188.5\arcsec$ on the $y$-axis represent the southwest and northeast endpoints of S2, respectively.}
\label{fig9}
\end{figure}

As mentioned before, the coronal loops to the North of the prominence are disturbed and start oscillating during the eruption 
(see animation in the Electronic Supplementary Material \textsf{animaia.mp4}).
In Figure~\ref{fig3}d1, two slices (S2 and S3) are used to investigate the transverse oscillations. 
The slightly curved slice S2 with a total length of $\approx$188.5$\arcsec$ is parallel to the solar surface. The straight slice S3 with a length of $\approx$172$\arcsec$ is vertical to the solar surface.
The time--distance diagram of S2 in 171\,{\AA} is displayed in Figure~\ref{fig9}. The positions of the oscillating loop with maximal EUV intensity are denoted by the magenta-plus symbols.
After the flare occurs, the coronal loop first moves southward and sways from side to side periodically. The amplitude decays with time and the oscillation lasts for three to four cycles.
The initial southward movement is caused by the strong magnetic-pressure gradient after the flux rope quickly escapes the active region.

\begin{figure} 
\centerline{\includegraphics[width=0.8\textwidth,clip=]{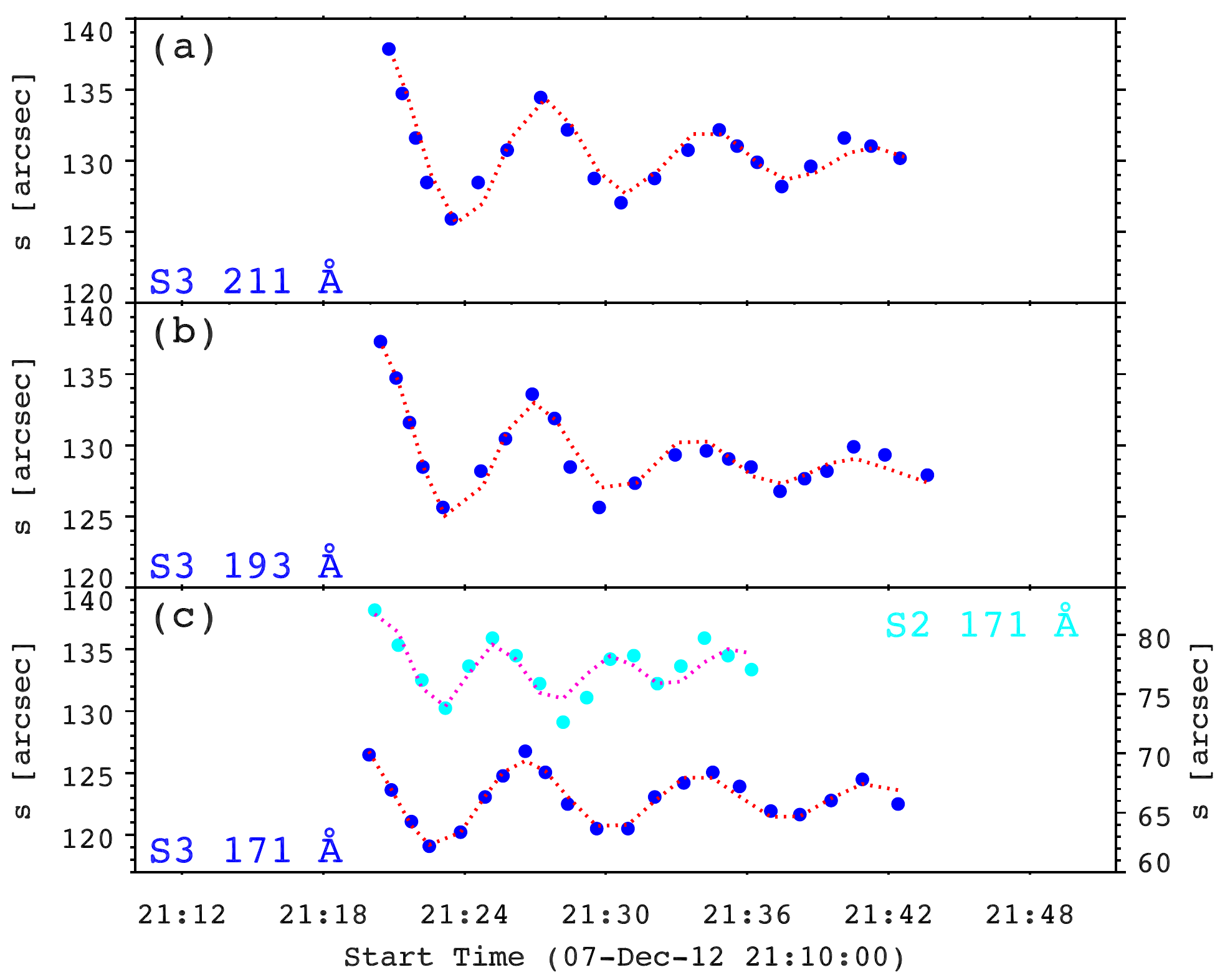}}
\caption{The positions (\textit{blue circles}) of the coronal loop tops along S3 in 171, 193, and 211\,{\AA}.
The results of curve fitting using Equation~\ref{eqn-2} are overlaid with \textit{red-dotted lines}.
In Panel {\bf c}, the positions of the coronal loop along S2 in 171\,{\AA} are plotted with \textit{cyan circles}.
The results of curve fitting using Equation~\ref{eqn-1} are overlaid with a \textit{magenta-dotted line}.}
\label{fig10}
\end{figure}

In Figure~\ref{fig10}c, the positions of the oscillating loop along S2 are drawn with cyan circles.
To precisely determine the physical parameters of the oscillation, we perform curve fitting by adopting an exponentially decaying sine function \citep{nis13}:
\begin{equation} \label{eqn-1}
  y(t)=A_{0}\sin(\frac{2\pi}{P}(t-t_0)+\phi_0)\mathrm{e}^{-(t-t_0)/\tau}+y_0+k(t-t_0)+c(t-t_0)^2,
\end{equation}
where $A_{0}$ and $\phi_0$ stand for the initial amplitude of displacement and phase at $t_0$, $P$ and $\tau$ signify the period and damping time,
$y_0$ denotes the initial loop position, and $k$ and $c$ denote the coefficients of linear and quadratic terms, respectively.
The results of curve fitting using \textsf{mpfit.pro} are drawn with a magenta-dotted line. 
It is evident that the transverse oscillation can be nicely described by Equation~\ref{eqn-1}.
In Table~\ref{tab-2}, the fitted values of $t_0$ ($\approx$21:20:11 UT), $A_{0}$ ($\approx$3.1\,Mm), $P$ ($\approx$294 s), $\tau$ ($\approx$645 s), 
and $\frac{\tau}{P}$ ($\approx$2.2) are listed in the second to sixth columns.
It should be emphasized that the transverse oscillation in the horizontal direction is most striking in 171\,{\AA} and becomes blurred in 193 and 211\,{\AA}.

\begin{table}
\caption{Parameters of the transverse coronal-loop oscillations in 171, 193, and 211\,{\AA}.
$A_{0}$ is the initial amplitude at $t_0$. $P$ and $\tau$ signify the period and damping time.
$h$ is the apparent height of loop top. $L$ is the loop length assuming a semi-circular shape.
$C_\mathrm{k}$ and $C_\mathrm{A}$ represent the phase speed and internal Alfv\'{e}n speed of the loop.
$B$ denotes the magnetic field strength of the loops.}
\label{tab-2}
\tabcolsep 1.5mm
\begin{tabular}{ccccccccccc}
\hline
slice & $t_0$ & $A_0$ & $P$ & $\tau$ & $\frac{\tau}{P}$ & $h$ & 2$L$ & $C_\mathrm{k}$ & $C_\mathrm{A}$ & $B$  \\
        & [UT]   &  [Mm]  &  [s]  &  [s]      &                          & [Mm] & [Mm] & [km\,s$^{-1}$] & [km\,s$^{-1}$] & [G] \\
  \hline
S2\_171 & 21:20:11 & 3.1$\pm$0.3 & 293.6$\pm$5 & 645.3$\pm$39 & 2.2 &  ... & ... & ... & ... & ... \\
S3\_171 & 21:19:47 & 3.4$\pm$0.3 & 441.4$\pm$8 & 1011.8$\pm$96 & 2.3 & 93.5 & 587.6 & 1331.3 & 987.6 & 12-37 \\
S3\_193 & 21:20:18 & 5.2$\pm$0.4 & 407.2$\pm$5 & 570.8$\pm$34 & 1.4 & 100.1 & 628.6 & 1543.8 & 1145.3 & 14-43 \\
S3\_211 & 21:20:47 & 5.0$\pm$0.4 & 419.2$\pm$6 & 641.4$\pm$43 & 1.5 & 100.1 & 628.6 & 1499.6 & 1112.5 & 13-41 \\
\hline
\end{tabular}
\end{table}

\begin{figure}
\centerline{\includegraphics[width=0.8\textwidth,clip=]{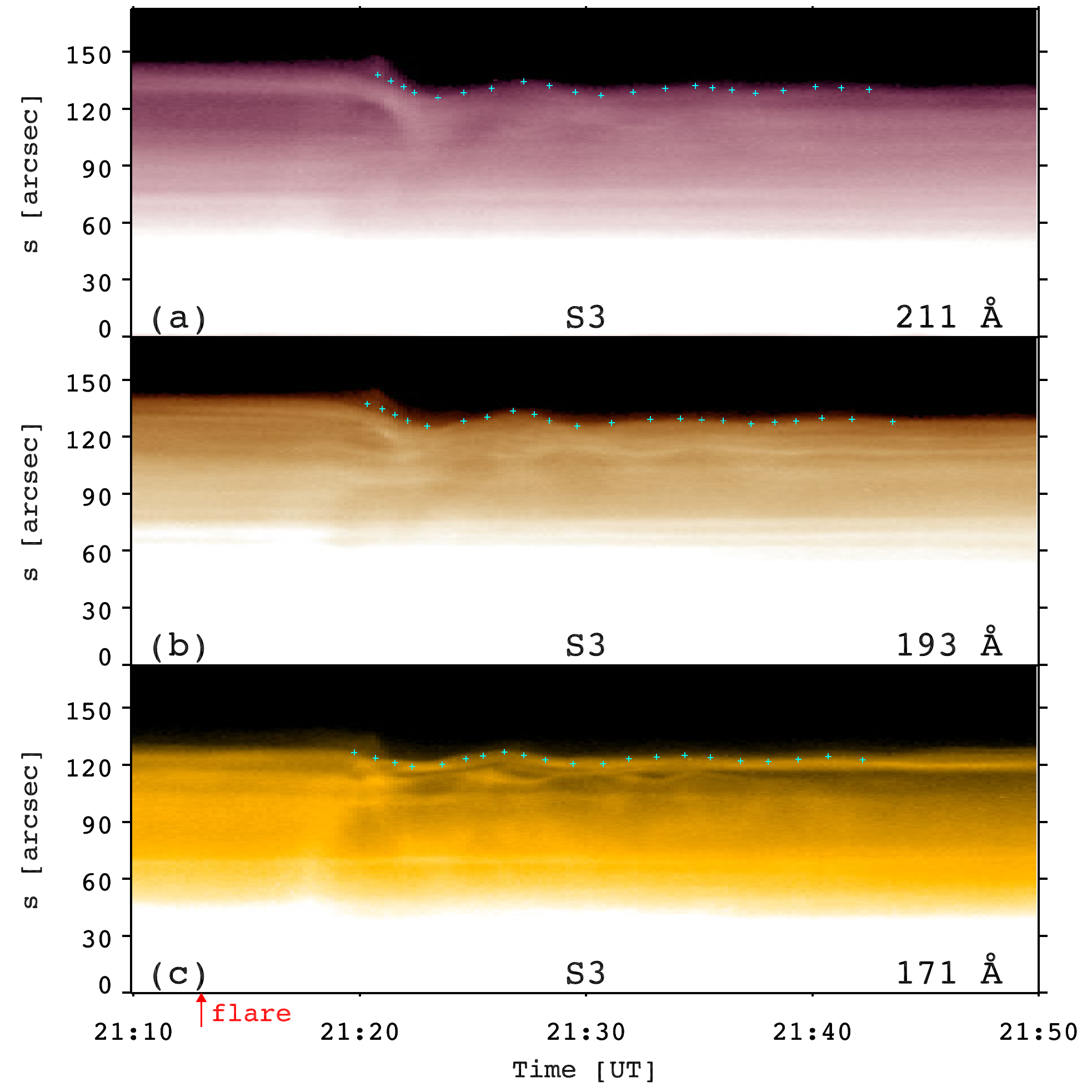}}
\caption{Time--distance diagrams of S3 in 171, 193, and 211\,{\AA}. The \textit{cyan-plus symbols} denote the positions of the loop tops.
The \textit{red arrow} on the $x$-axis indicates the starting time of flare. $s=0$ and $s=172\arcsec$ on the $y$-axis represent the east and west endpoints of S3.}
\label{fig11}
\end{figure}

The time--distance diagrams of S3 in 171, 193, and 211\,{\AA} are displayed in Figure~\ref{fig11}. The positions of the loop tops are denoted by the cyan-plus symbols.
Interestingly, the coronal loops start to oscillate vertically after the flare occurs. The amplitudes also decay with time and the oscillations last for three to four cycles.
The initial inward motion indicates loop contraction or implosion before oscillation \citep[e.g.][]{gos12,sun12a,sim13,dud16}. 
This is consistent with the fact that the vertical loop oscillations are caused by the eruption of the flux rope, since the magnetic pressure beneath the loops is impulsively decreased after the eruption.

The positions of the loop tops in 171, 193, and 211\,{\AA} are manually determined and are drawn with blue circles in Figure~\ref{fig10}.
Likewise, we perform curve fitting by adopting an exponentially decaying sine function:
\begin{equation} \label{eqn-2}
  y(t)=A_{0}\sin(\frac{2\pi}{P}(t-t_0)+\phi_0)\mathrm{e}^{-(t-t_0)/\tau}+y_0+k(t-t_0),
\end{equation}
where the parameters have the same meanings as in Equation~\ref{eqn-1}.
The results of curve fitting are drawn with red-dotted lines, showing that the vertical oscillations can be well described by Equation~\ref{eqn-2}.
In Table~\ref{tab-2}, the fitted values of $t_0$, $A_{0}$, $P$, $\tau$, and $\frac{\tau}{P}$ are listed in the second to sixth columns.
The initial amplitudes lie in the range of 3.4\,--\,5.2\,Mm with an average value of $\approx$4.5\,Mm. The periods have a range of 407\,--\,441 s with an average value of $\approx$423 s.
The damping times are between 570 s and 1012 s with an average value of $\approx$741 s. 
The corresponding quality factors ($\frac{\tau}{P}$) are between 1.4 and 2.3 with a mean value of $\approx$1.7, suggesting a quick attenuation \citep{wht12b}.
In 171\,{\AA}, the period of oscillation in the vertical direction is $\approx$1.5 times longer than that in the horizontal direction, while the quality factors are close to each other,
indicating that the transverse oscillations in both directions are independent rather than two components of the same oscillation.
In Table~\ref{tab-2}, the apparent heights of the loop tops are listed in the seventh column.
The loop heights in 193 and 211\,{\AA} are higher than those in 171\,{\AA} with lower temperature of the response function peak \citep{lem12}. 
In other words, the coronal loops with different heights oscillate in phase \citep{zqm20b}.

The commencements of oscillations in both horizontal and vertical directions occur during the fast rise of the flux rope (Figure~\ref{fig7}a), 
which is in accordance with the fact that the transverse oscillations are caused by the flux-rope eruption.
That is to say, the coronal loops are not disturbed until the flux rope expands and propagates after a time.
It is emphasized that the simultaneous oscillations in the horizontal direction and vertical directions are from different coronal loops superposed along the LOS,
since the coronal loop undergoing horizontal oscillation drifts southward from the equilibrium position (see Figure~\ref{fig9}).
Moreover, the period of vertical oscillation is 1.5 times longer than the horizontal oscillation in 171\,{\AA}.

\section{Discussion} \label{s-dis}
\subsection{What is the Cause of Non-radial Eruption?} \label{s-nr}
\begin{figure} 
\centerline{\includegraphics[width=0.9\textwidth,clip=]{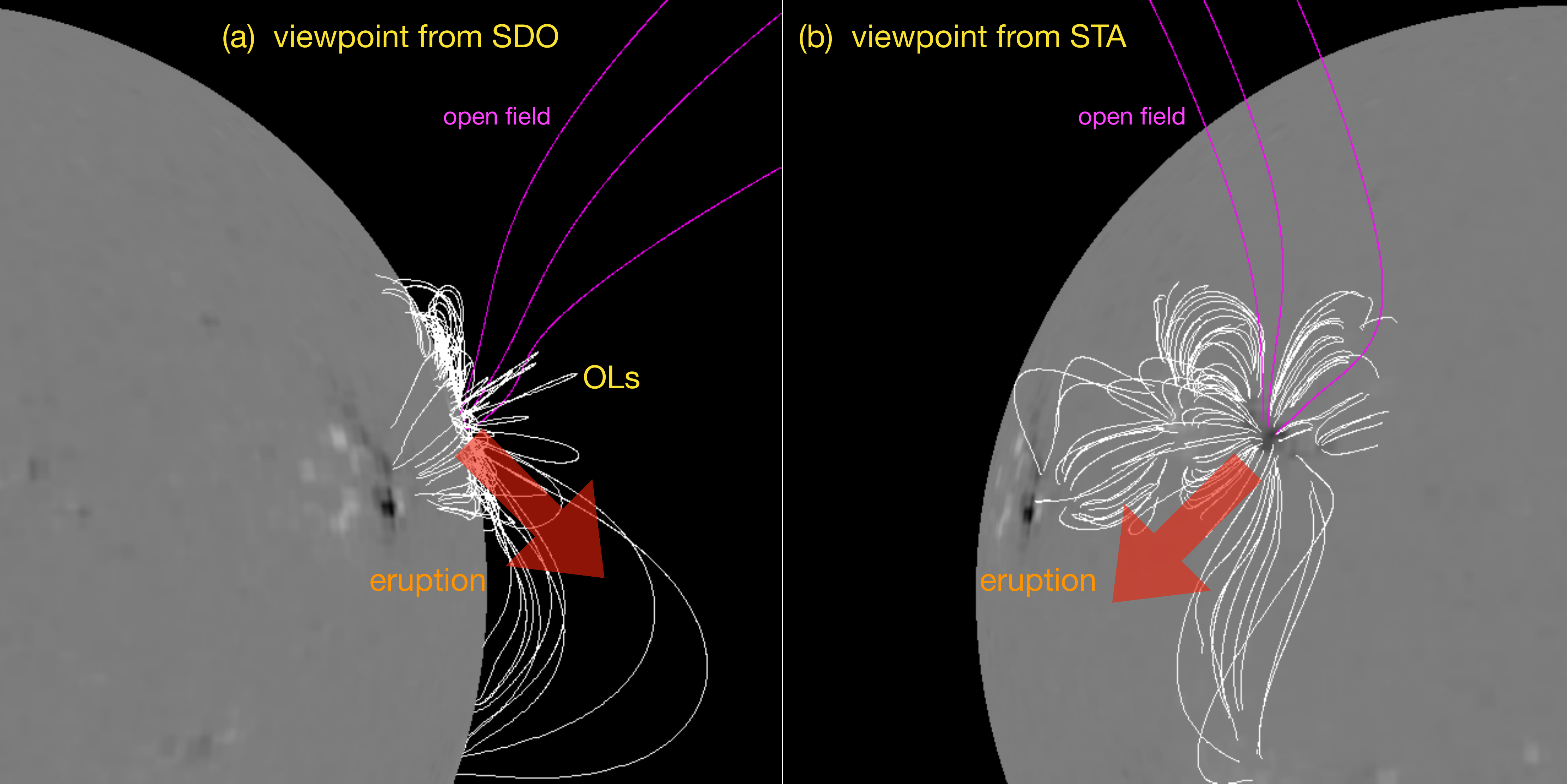}}
\caption{The 3D magnetic configuration at 18:04:00 UT from the viewpoints of SDO ({\bf a}) and STA ({\bf b}) using the PFSS modeling.
The \textit{white and magenta lines} represent the closed and open field, respectively. The \textit{red arrows} indicate the directions of prominence eruption.}
\label{fig12}
\end{figure}

As mentioned in Section~\ref{s-intro}, non-radial prominence eruptions are frequently observed \citep[e.g.][]{wil05,gos09,sun12b,bi13,pan13,yang18,mi19,devi21,guo21b,man21}.
The reported inclination angle is between 45$^\circ$ and 90$^\circ$. The non-radial eruption is attributed to the imbalance of magnetic pressure of the high-lying field \citep{au10}.
In the current case, the apparent inclination angle of the flux rope is 60$^\circ$. We obtain the global 3D magnetic configuration at 18:04:00 UT 
using the potential-field source surface \citep[PFSS:][]{sch03} modeling. Figure~\ref{fig12} shows the configuration from the viewpoints of SDO (a) and STA (b), respectively.
Combining Figure~\ref{fig12} and Figure~\ref{fig3}, it is revealed that the flux rope finds a way out where the magnetic field is weaker and the escape becomes easier. 
This is consistent with the previous interpretation \citep{au10}.

\subsection{How are the Transverse Loop Oscillations Excited?} \label{s-trigger}
Two decades have passed since the discovery of coronal-loop oscillations \citep[see][and references therein]{li20,naka20,naka21,wang21}.
Kink oscillation has become a topic of great interest due to its advantage in diagnosing the coronal magnetic field \citep{yang20}. 
There are several candidates in exciting kink oscillations, such as flare-induced blast waves \citep{naka99,zqm20a}, large-scale coronal waves \citep{kum13}, coronal jets \citep{dai21}, 
lower coronal eruptions/ejections \citep[LCE:][]{zim15}, coronal rains \citep{ant11}, and reconnection outflows \citep{ree20}. 
A common characteristic of the above excitation mechanisms is that the coronal loops start oscillating after being impacted. 
In our case, a pressure depletion is created as a result of the non-radial flux-rope eruption. 
Hence, the transverse loop oscillations are driven by the strong magnetic-pressure gradient of the loops (see \textsf{animaia.mp4}), which is rarely noticed and reported.
Numerical simulations are desired to justify this mechanism.

In Figure~\ref{fig9} and Figure~\ref{fig11}, there are several oscillating threads with non-zero phase differences compared to the analyzed loops. 
This is probably because a bundle of loops with different lengths and periods oscillate simultaneously \citep{wht12a,nis13}. We focus on oscillations with clear and complete signals.
It is emphasized that our analysis has LOS limitations in classifying oscillations into horizontal and vertical types based on single-point observation.
The possibility of elliptically polarized transverse oscillations that are decomposed into two linearly polarized modes with different periods could not be excluded.
Forward modeling and multi-point observations are required to clarify the polarization of kink oscillations \citep{wht12b}.

\subsection{Magnetic Field Estimated from Coronal Seismology} \label{s-mag}
To estimate the magnetic field of the loops undergoing vertical oscillations, we use the observed periods and coronal seismology. 
The phase speed [$C_\mathrm{k}$] of the standing kink oscillation is determined by the loop length [$L$] and period \citep{naka99}:
\begin{equation} \label{eqn-3}
C_\mathrm{k}=\frac{2L}{P}=\sqrt{\frac{2}{1+\rho_\mathrm{e}/\rho_\mathrm{i}}}C_\mathrm{A},
\end{equation}
where $C_\mathrm{A}$ is the internal Alfv\'{e}n speed, and $\rho_\mathrm{e}$ and $\rho_\mathrm{i}$ represent the external and internal plasma densities. 
The lengths of the oscillating loops are listed in the eighth column of Table~\ref{tab-2} based on a semi-circular shape.
The corresponding values of $C_\mathrm{k}$ and $C_\mathrm{A}$ are listed in the ninth and tenth columns, assuming that $\rho_\mathrm{e}/\rho_\mathrm{i}=0.1$ \citep{naka01}.

The magnetic-field strength of the oscillating loops is determined by $\rho_\mathrm{i}$ and $C_\mathrm{A}$, $B=\sqrt{4\pi\rho_\mathrm{i}}C_\mathrm{A}$.
Since the LOS depths of the oscillating loops are difficult to measure, we could not determine $\rho_\mathrm{i}$ precisely.
Assuming that $\rho_\mathrm{i}$ is between 1.1$\times$10$^{-15}$\,g\,cm$^{-3}$ and 1.1$\times$10$^{-14}$\,g\,cm$^{-3}$,
corresponding to the electron number density between 0.7$\times$10$^{9}$\,cm$^{-3}$ and 7$\times$10$^{9}$\,cm$^{-3}$ \citep[e.g.][]{naka01,van08,ver09,ver13,yuan16,dai21},
the magnetic-field strengths of the loops observed in different wavelengths are calculated and listed in the last column of Table~\ref{tab-2}. 
Hence, the magnetic fields of the vertically oscillating loops fall in the range of 12\,--\,43\,G.
It is noted that the estimated magnetic field using coronal seismology has large uncertainties.
On one hand, the plasma density could not be precisely determined, so that a wide range is adopted according to previous literature.
On the other hand, the shape of the loops is unknown, so that a simple semi-circular shape is employed.
Besides, the apparent heights [$h$] and corresponding lengths [$L$] of the loops are lower limits of the true values.
The estimations of $C_\mathrm{k}$, $C_\mathrm{A}$, and $B$ in Table~\ref{tab-2} are lower limits as well.
Considering that the loops undergoing horizontal and vertical oscillations are heavily superposed along the LOS, the height of oscillating loop in the horizontal direction is hard to determine.
Therefore, the horizontal oscillation is not used for seismology.

\section{Summary} \label{s-sum}
In this article, we investigate the transverse coronal loop oscillations induced by the eruption of a prominence-carrying flux rope on 7 December 2012.
The main results are as follows:
\begin{enumerate}[(i)]
\item The flux rope originating from AR 11621 is observed in various EUV wavelengths, suggesting its multi-thermal nature.
The early evolution of the flux rope is divided into two phases: a slow rise phase at a speed of $\approx$230\,km\,s$^{-1}$ and a fast rise phase at a speed of $\approx$706\,km\,s$^{-1}$.
The eruption generates a C5.8 flare and the onset of fast rise is consistent with the HXR peak time of the flare. The embedded prominence has a lower speed of $\approx$452\,km\,s$^{-1}$.
The propagation of the flux rope is along the southwest direction in the FOV of AIA. 
Hence, the inclination angle between the direction of flux rope eruption and the local solar normal reaches $\approx$60$^{\circ}$, suggesting a typical non-radial eruption.
\item During the early eruption of the flux rope, the nearby coronal loops are disturbed and experience kink-mode oscillations.
The oscillation in the horizontal direction has an initial amplitude of $\approx$3.1\,Mm, a period of $\approx$294 s, and a damping time of $\approx$645 s. 
It is most striking in 171\,{\AA} and lasts for three to four cycles. The oscillations in the vertical directions are observed mainly in 171, 193, and 211\,{\AA}.
The initial amplitudes lie in the range of 3.4\,--\,5.2\,Mm, with an average value of 4.5\,Mm. 
The periods are between 407\,seconds and 441\,seconds, with an average value of 423\,seconds ($\approx$7 minutes).
The oscillations are damping and last for nearly four cycles. The damping times lie in the range of 570\,--\,1012\,seconds, with an average value of 741\,seconds.
\item Assuming a semi-circular shape of the vertically oscillating loops, we calculate the loop lengths according to their heights. 
Using the observed periods, we carry out coronal seismology and estimate the internal Alfv\'{e}n speeds (988\,--\,1145\,km\,s$^{-1}$) and the magnetic-field strengths (12\,--\,43\,G) of the oscillating loops.
\end{enumerate}

\begin{acks}
The authors thank the reviewer for valuable comments and suggestions.
We thank L. Feng and Y.N. Su in Purple Mountain Observatory for helpful discussions.
SDO is a mission of NASA\rq{}s Living With a Star Program. AIA and HMI data are courtesy of the NASA/SDO science teams.
This work is funded by the National Key R\&D Program of China 2021YFA1600502 (2021YFA1600500), NSFC grants (No. 11790302, 11773079, 11973012, 11973092, 12073081), 
the International Cooperation and Interchange Program (11961131002), CAS Key Laboratory of Solar Activity, National Astronomical Observatories (KLSA202113),
the Strategic Priority Research Program on Space Science, CAS (XDA15052200, XDA15320301), and the mobility program (M-0068) of the Sino-German Science Center.
\end{acks}

\section*{Declarations}
\textbf{Disclosure of Potential Conflicts of Interest} The authors declare that they have no conflicts of interest.

\bibliographystyle{spr-mp-sola}
\bibliography{SOLA21-147_Zhang}

\end{article}
\end{document}